\documentclass[12pt]{article}
\usepackage{epsfig}
\begin{document}
\title{Front propagation in random media:\\
From extremal to activated dynamics}
\author{Rune Skoe, Damien Vandembroucq and St\'ephane Roux\\
Unit\'e Mixte CNRS/Saint-Gobain ``Surface du Verre et Interfaces'',\\ 
39 Quai Lucien Lefranc, 93303 Aubervilliers Cedex, France}
\maketitle
\begin{abstract} 
Front propagation in a random environment is studied close to the
 depinning threshold. At zero temperature we show that the depinning
 force distribution exhibits a universal behavior. This property is
 used to estimate the velocity of the front at very low
 temperature. We obtain a Arrhenius behavior with a prefactor
 depending on the temperature as a power law. These results are
 supported by numerical simulations.
\end{abstract}

\section{Introduction}

Front propagation in a random environment has in recent years
become a generic problem describing the motion of
interfaces as different as magnetic domain walls\cite{LemerlePRL98},
wetting\cite{Rolley} or fracture fronts\cite{EBouchaud97}. In all
cases the balance between a roughening effect due to the quenched
random environment and the smoothing effect of the elastic
interactions leads to a rich phenomenology. At zero temperature one
observes a depinning transition: a threshold force is required for unlimited motion.
When the front is driven at a force
close to the threshold, one observes spatio-temporal organization, the
front roughness exhibits self-affine correlations and the critical
exponents characterising the propagation of the front can be shown to
depend on the nature of the elastic interactions\cite{TGRPRE98}.  At
finite temperature the dynamics is controlled by thermal activation
and a creep regime characterized by a stretched exponential dependence
of the velocity on the driving force has been predicted by
scaling\cite{creepscaling} or renormalisation \cite{ChauvePRB00}
arguments. This creep regime has been experimentally observed in weak
pinning conditions for the propagation of a magnetic domain wall
\cite{LemerlePRL98}.

In the following we consider the case of strong pinning, we first give 
results obtained in the framework of an extremal dynamics focusing on
the depinning force distribution in the vicinity of the macroscopic
depinning threshold. We  then consider the thermally activated
propagation of a front driven at a constant force and we show that at
low temperature the knowledge of the depinning force distribution
leads to a simple Arrhenius regime with a prefactor that depends on
the temperature as a power law. To illustrate these results we present
in the following numerical simulations performed with two different
elastic interactions: Laplacian (corresponding to the motion of
magnetics walls or fluid invasion in a Hele-Shaw cell) and long range
elasticity (which corresponds to  the propagation
of a wetting or a fracture front).

\section{Force distribution in the extremal model}  

We consider the propagation of an elastic 1D front through a 2D random
environment. The position of the front at abscissa $x_i$ is given by
$h_i=h(x_i)$. The random environment consists in traps of depth
$\gamma_i$ which are randomly distributed in the direction of
propagation. In the following we restrict ourselves to the case of an
overdamped dynamics in a strong pinning situation. When submitted to
an external driving force $f_{ext}$ the depinning criterion for the
site $i$ can be written as:

\begin{equation}
f_{ext} +f_{el}[x_i,h_i] > \gamma_i
\end{equation}
where the elastic component can be Laplacian or long range. In the
particular case of an interfacial fracture front, the long range
elastic kernel comes from a first order expansion of the stress
intensity factor\cite{GaoRice89} and with periodic boundary conditions
it can be written:
\begin{equation}
f_{el}[x_i,h_i] = 
\sum_{j\ne i} \frac{h_j-h_i}{\sin\left[\pi(x_i-x_j)/L\right]^2}\;,
\end{equation}
where $L$ is the size of the system. In the Laplacian case, we use a
discrete form of the Laplace operator:
\begin{equation}
f_{Lap} [x_i,h_i] = \frac{1}{2} \left(h_{i+1}- 2h_i + h_{i-1}\right)
\end{equation}

In the extremal dynamics the external force is adapted at every step
to the value $f_{ext}=f_c(t)$ such that only one site can depin. This
corresponds to selecting the weakest site $i_0$ of the front:
\begin{equation}
f_c=\gamma_{i_0} - f_{el}\left[x_{i_0},h_{i_0}\right]=min_i \left( \gamma_i - f_{el}\left[x_i,h_i\right] \right)\;.
\end{equation}
The weakest site is then advanced up to the next trap, the elastic
forces are updated and the process is iterated on the weakest site of
the new configuration. The maximum of these depinning forces $f_c$
over time corresponds to the macroscopic depinning threshold $f^*$.

The front then presents a multiscale roughness: it remains
statistically invariant under the anisotropic scaling transformations
$x\to \lambda x$, $y \to \lambda^{\zeta} y$. The front is said to be
self-affine and the exponent $\zeta$ is called the roughness
exponent. This scaling invariance is responsible for long range
correlations of the height differences on the front. In particular
when measured over a size $d$, the height standard deviation grows as
a power law: $\sqrt{<\Delta y^2>}(d) \propto d^{\zeta}$. A direct
consequence of this self-affinity property is that the height power
spectrum is a power law of exponent $-1-2\zeta$ where $\zeta \simeq
0.35$ and $\zeta \simeq 1.25$ in the fracture front case and in the
Laplacian case respectively.

It can be shown that the behavior of the ditribution ${\cal Q}(f_c)$
in the vicinity of the threshold $f^*$ exhibits universal features. In
order to illustrate this point let us focus on a sequence of
successive depinning events on the front. Such sequences are known to
be time and space correlated and present an avalanche-like
behavior\cite{TGRPRE98}. The distance $d$ between two successive
events is distributed according to $p(d)\propto d^{-a}$ where $a=3$
 in the Laplacian case and $a=2$, the exponent of the
elastic kernel, in the case of long range elastic interactions. Large
separations between two succesive events correspond to the jump from
one avalanche to another one, the larger the jump, the closer the
depinning force to the threshold $f^*$. Let us consider the force
distribution $q_d(f)$ conditionned to the size $d$ of the jump. From
the knowledge of the elastic interactions and of the front roughness
statistics we can give estimates of the moments of these
distributions. The typical height fluctuations on a distance $d$
scaling as $d^{\zeta}$, the force fluctuations scale as $d^{-b}$ where
$b={1-\zeta}\simeq 0.65$ in the case of long range interactions or
$b={2-\zeta}\simeq 0.75$ in the Laplacian case. As $d$ increases, 
$q_d(f)$ becomes narrower and closer to the threshold $f^*$. In
particular we get the linear relationship:
\begin{equation}
\label{f*}
\langle f_c \rangle_d = f^* -A \sigma_d(f_c)\;,
\end{equation}
where $\langle f_c \rangle_d$ and $\sigma_d(f_c)$ are the mean and the
standard deviation of the depinning forces conditionned to a jump of
size $d$. This relationship allows a precise determination of the
threshold $f^*$. Moreover all distributions $q_d$ are identical up to a
rescaling transformation:
\begin{equation}
\label{force-scaling}
q_d\left( f^*-f_c\right) = d^b \psi \left[ (f^*-f_c)d^b \right]\;.
\end{equation}
The knowledge of the distribution $p(d)\propto d^{-a}$ of the
distances between successive active sites allows to express the
depinning force distribution close to the threshold:
\begin{equation}
\label{universal}
{\cal Q}(f^*-f_c) =\int x^{b-a} \psi \left[ (f^*-f_c)x^b \right] dx
 \propto \left(f^*-f_c \right)^{\nu} \;, \quad \nu={\frac{a-b-1}{b}} \;,
\end{equation}
vith the numerical values $\nu \simeq 0.5$ in the fracture front case
and $\nu \simeq 1.7$ in the Laplacian case.
\begin{figure}
\label{force-distribution}
\epsfig{file=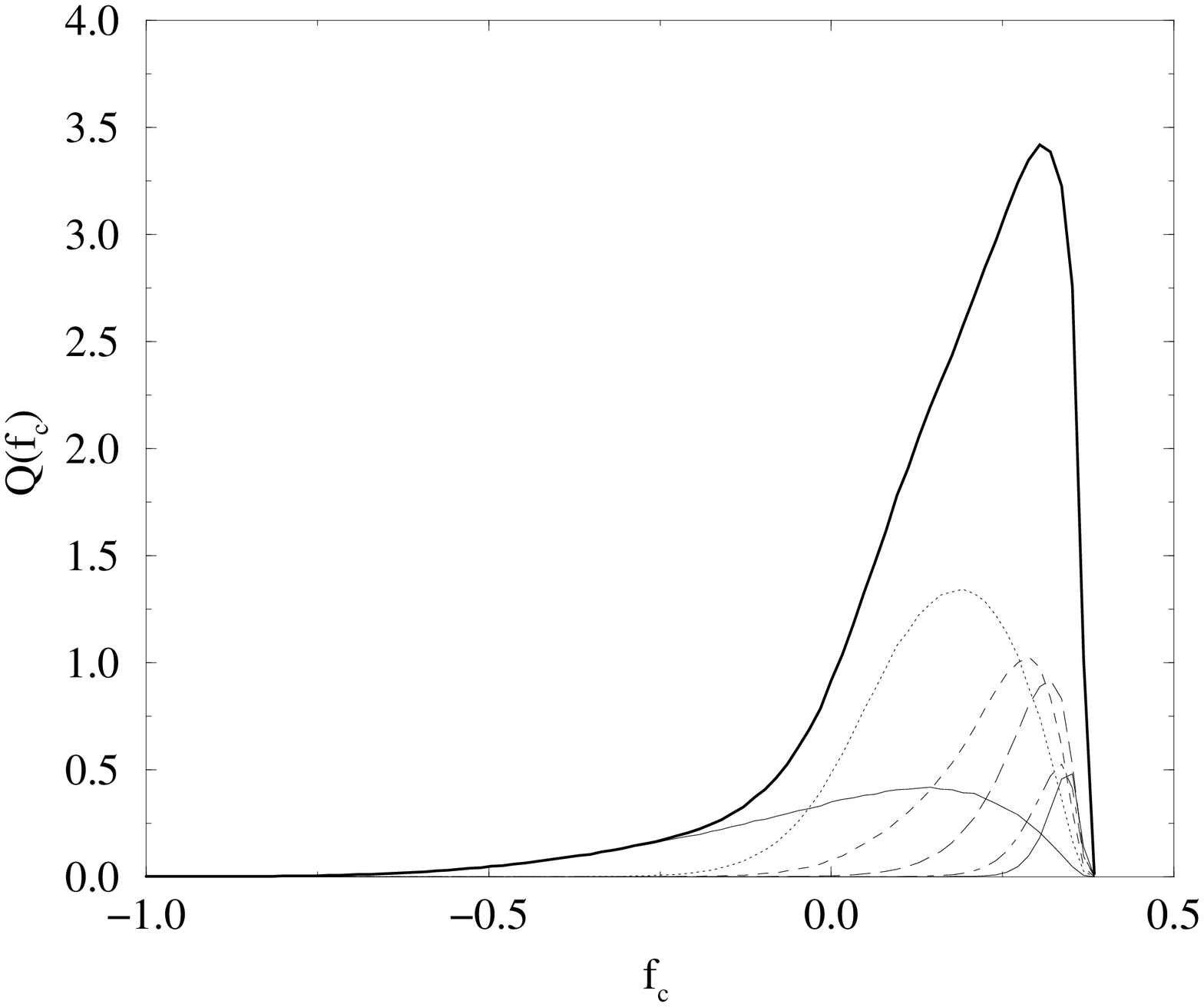,width=0.42\textwidth,angle=-90}
\hfill
\epsfig{file=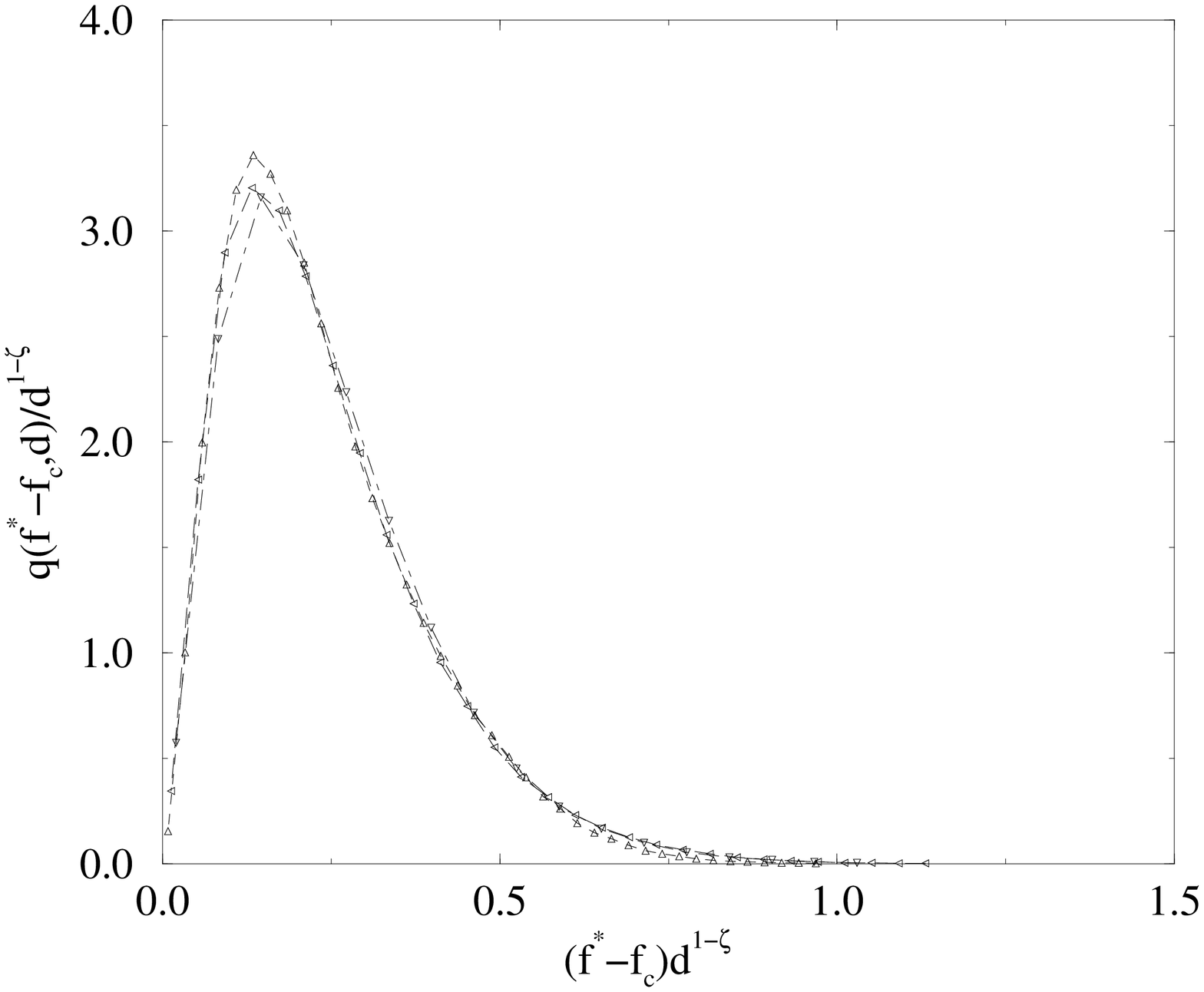,width=0.42\textwidth,angle=-90}
\caption{Left: Distribution of depinning forces (bold) and
contributions conditionned by the distances between consecutive active
sites. Apart from the first distribution (continuous line) associated
with very short jumps, which is sensitive to the details of the random
threshold distribution, the other ones present the same tendency: the
larger the jump, the closer the mean force to the threshold and the
narrower the distribution.  Right: these distributions are
scaling invariant, after rescaling according to
Eq. (\ref{force-scaling}) all these distribution fall on the same
master curve. These results have been obtained in the case of long
range elastic interactions.}
\end{figure}

In Fig. 1 we present numerical results obtained in the case of long
range elastic interactions, we see that the low tail contribution
corresponds to activity jumps of size 0 or 1. For contributions
corresponding to larger jumps, the details of the random trap
distribution are washed out and we can check that the force
distributions obey the scaling proposed in
Eq. (\ref{force-scaling}). Note in addition that the universal
behavior proposed in Eq. (\ref{universal}) only applies in a very
close vicinity of the threshold $f^*$.  The size of the sytem is
$L=1024$, the trap depth and the distance between two traps in the
propagation direction are uniformly distributed in the range
$[0-1]$. The results are averaged over 5 millions iteration steps. A
linear fit using Eq. (\ref{f*}) gives $f^*=0.373\pm0.001$ and we get
for the roughness exponent $\zeta=0.34\pm0.01$.

\section{From extremal to activated dynamics}

Instead of adapting the driving force at every step so that only one
site can depin at a time we consider now the front propagation under a
constant external driving force $f_{ext}$. At zero temperature the
front moves freely if $f_{ext} > f^*$, while it will only advance a
finite distance before being pinned if $f_{ext} < f^*$.  This
travelling distance diverges when $f_{ext}$ approaches the threshold
$f^*$. At finite temperature we now allow a pinned site $i$ to depin
at any time with an Arrhenius probability $p_i$:

\begin{equation}
p_i=e^{\frac{\Delta_i}{T}}\;,\quad \Delta_i=f_{ext}+f^{el}_i-\gamma_i<0\;,
\end{equation}
and (assuming all other sites to be frozen) the probability that the
site has remained pinned during the time $\tau$ and advances in the
interval $[\tau,\tau+d\tau]$ can be written

\begin{eqnarray}
\displaystyle{
P_i(\tau)d\tau =\frac{1}{\tau_i}e^{-\frac{\tau}{\tau_i}}  d\tau\;,\quad 
\tau_i=\frac{1}{p_i}=e^{-\frac{\Delta_i}{T}}\;.
}
\end{eqnarray}
If we consider now all sites of the front,
 the typical waiting time for the first depinning event on the
front is $\tau^*$ such that:
\begin{equation}
\frac{1}{\tau^*}=\sum_i \frac{1}{\tau_i}=\sum_i e^{\frac{\Delta_i}{T}}\;,
\end{equation}
the probability that the depinning takes place at site $i$ being
$p_i/\sum p_i=\tau^*/\tau_i$. At very low temperature the waiting time
$\tau^*$ is dominated by the waiting time of the weakest site $i_0$
\begin{equation}
\frac{1}{\tau^*}=\frac{1}{\tau_{i_0}}
 \left( 
1+\sum_{i\ne i_0} 
 e^{\frac{\Delta_i-\Delta_{max}}{T}}
 \right)\;.
\end{equation}

In the following we focus on this situation of very low temperatures.
We consider the case of an elastic front driven at a constant force
$f_{ext}$. The (very low) temperature allows to introduce a physical
time in the extremal model which is adapted as follows.  When the
driving force $f_{ext}$ exceeds the local threshold $f_c$, all sites
which fulfill the depinning criterion are allowed to advance at the
same time. Then the elastic forces are updated to take into account the change
of front conformation.  If however the site reaches a blocked
configuration where no site is able to depin at zero temperature, the
weakest site is allowed to advance by thermal activation. The waiting
time $t_0$ is chosen according to the distribution $(1/\tau_{i_0})
\exp(-t_0/\tau_{i_0})$ corresponding to the extremal site and the
elastic forces are updated before the next iteration.

The distribution of the depinning forces being known in the vicinity
of the macroscopic threshold $f^*$, we can estimate the creep velocity
at very low temperature {\it via} the average waiting time of an
activated step:

\begin{eqnarray}
\label{integral}
\langle \tau \rangle 
&\simeq & T^{1+\nu}\exp \left( \frac{f^*-f_{ext}}{T}  \right)
 \int_0^{\frac{f^*-f_{ext}}{T}}  u^{\nu} e^{-u}du \;,
\end{eqnarray}
so that we expect at low temperatures:
\begin{equation}
v(f_{ext},T) = T^{-(1+\nu)} \exp \left(- \frac{f^*-f_{ext}}{T}  \right)\;.
\end{equation}
and a simple Arrhenius behavior when $T \gg f^*-f_{ext}$.

\begin{figure}
\label{creep}
\epsfig{file=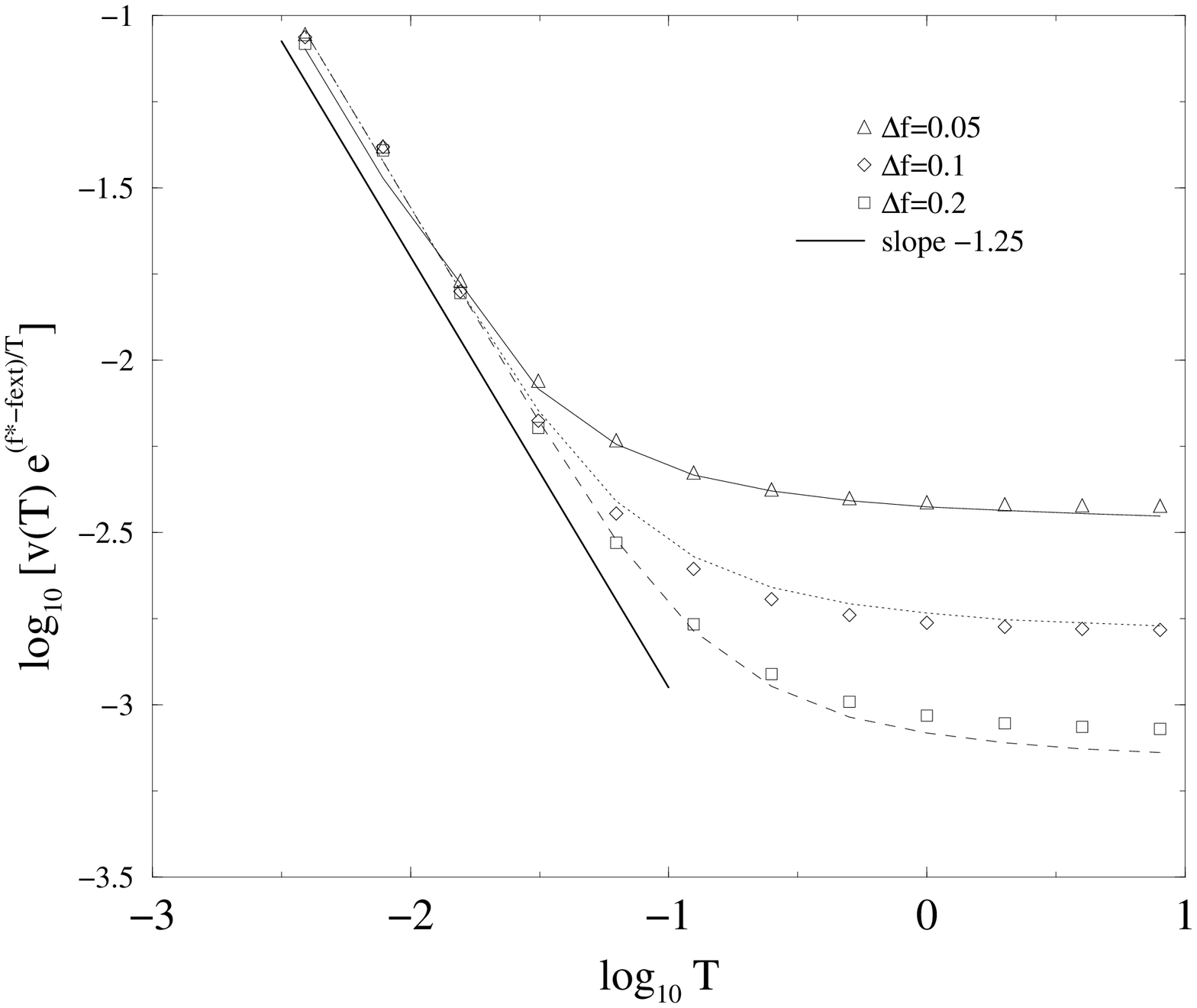,width=0.42\textwidth,angle=-90}
\hfill
\epsfig{file=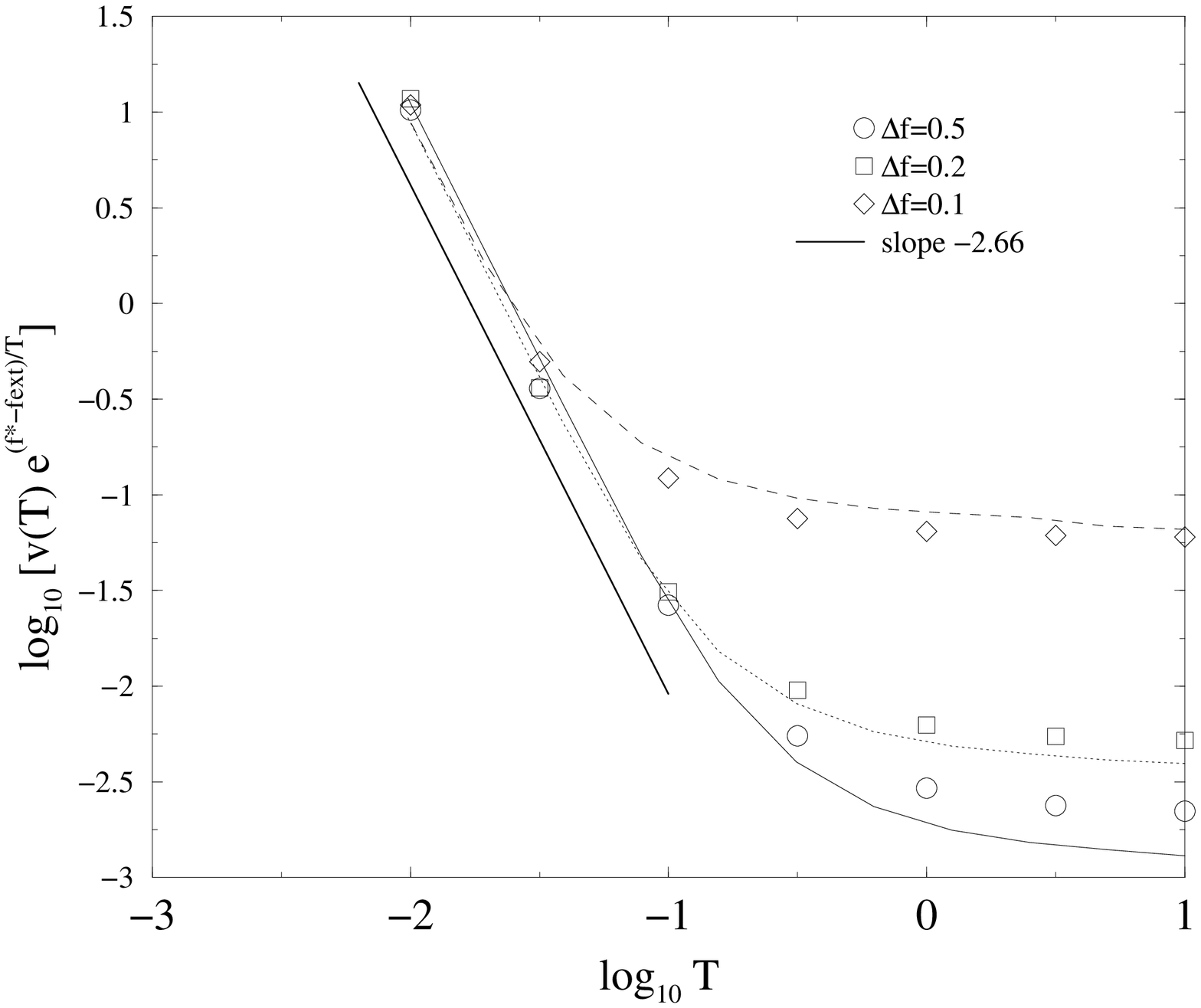,width=0.42\textwidth,angle=-90}
\caption{Velocity of the front after rescaling by the Arrhenius term
$\exp[(f^*-f_{ext})/T]$ for different driving forces in the case of
long range elastic interactions (left) in Laplacian elasticity
(right). The symbols correspond to the numerical simulations and the
lines to the expected results from Eq. (\ref{integral} with an
exponent $\nu=0.25$ (long range) and $\nu=1.66$ (Laplacian)}
\end{figure}

In Fig. 2 we present the numerical results obtained for a
front driven at different forces below the threshold. The symbols
correspond to the velocity results after correction by the Arrhenius
term and the lines to the expected behavior following
Eq. \ref{integral}. For the latter expressions we used the values
$\nu=0.25$ in the long range case and $\nu=1.66$ and Laplacian case to
compare with the expected values $\nu=0.5$ and $\nu=1.7$
respectively. These differences may be due to the fact that the
universal behavior of the depinning force distribution is only valid
in a  region close to the threshold ($\Delta f < 0.1$  in our
case, see Fig. 1) and for lower driving forces the system becomes more
sensitive to the details of the distribution.

This behavior is different from the stretched exponential described in
\cite{creepscaling,ChauvePRB00,MarchettiPRB95,MarchettiPRL96}; the
main reason is probably that although we consider very low
temperatures, our study is restricted to the very close vicinity of
the threshold while the usual scaling obtained for creep motion
corresponds to driving forces far below the threshold. Note in
addition that at very low temperature, the time scales of depinning
become very large and that it may be necessary to take other
mechanisms into account. This is for example the case for crack
propagation in glass for which the subcritical motion of the front
competes with a stress assisted ion interdiffusion phenomenon which
tends to locally reinforce the material\cite{nghiemPhD98}. 
Within these restrictions, our knowledge of the depinning force
distribution gives us the  evolution of the front velocity with
temperature by way of the front roughness, thus relating a dynamical
behavior to a geometrical property.


\begin{thebibliography}{10}

\bibitem{LemerlePRL98}S. Lemerle, J. Ferr\'e, C. Chappert,
 V. Mathet, T. Giamarchi and P. Le Doussal,
Phys. Rev. Lett. {\bf 80}, 849, (1998) 

\bibitem{Rolley} E. Rolley, C Guthmann and R. Gombrowicz,
Phys. Rev. B {\bf 80}, 2865, (1998) 

\bibitem{EBouchaud97} E. Bouchaud,
J. Phys. Cond. Mat. {\bf 9}, 4319, (1997)

\bibitem{TGRPRE98} A. Tanguy, M. Gounelle and S. Roux, 
Phys. Rev. E {\bf 58}, 1577, (1998) 

\bibitem{creepscaling} L.B. Ioffe and V. M. Vinokur,
J. Phys. C.  {\bf 20}, 6149, (1987)

\bibitem{ChauvePRB00} P. Chauve, T. Giamarchi and P. Le Doussal,
Phys. Rev. B {\bf 62}, 6241, (2000) 


\bibitem{MarchettiPRB95}
L.-W. Chen and M.C. Marchetti,
Phys. Rev. B {\bf 851}, 6296, (1995) 

\bibitem{MarchettiPRL96}
V. M. Vinokur, M.C. Marchetti and L.-W. Chen,
Phys. Rev. Lett. {\bf 77}, 1845, (1996) 


\bibitem{GaoRice89} H. Gao and J.R. Rice,
J. Appl. Mech. {\bf 56}, 828, (1989) 

\bibitem{nghiemPhD98} B. Nghi\^em, 
Ph. D. Thesis, Universit\'e Paris VI, (1998)

\end{thebibliography}
\end{document}